# SiO$_x$/Si radial superlattices and microtube optical ring resonators

*R. Songmuang\*, A. Rastelli, S. Mendach, and O. G. Schmidt*
Max-Planck-Institut für Festkörperforschung, Heisenbergstr. 1, D-70569 Stuttgart, Germany

Scanning and transmission electron microscopy reveal that SiO$_x$/Si layers can roll-up into microtubes and radial superlattices on a Si substrate. These hybrid objects are thermally stable up to 850 °C and emit light in the visible spectral range at room temperature. For tubes disengaged from the substrate surface, optically resonant emission with mode spacings inversely proportional to the tube diameter are observed and agree excellently with those obtained from Finite-Different-Time-Domain simulations. The resonant modes we record are strictly polarized along the tube axis.

\*To whom correspondence should be addressed: R.Songmuang@fkf.mpg.de

Repetitive total light reflection inside circular geometries has been employed in several types of optical resonators such as microdisks[1], microtoroids[2], or microspheres[3]. Very recently, waveguiding and optical resonant modes have also been observed at cryogenic temperatures in rolled-up quantum dots in a tube.[4,5] Such tube structures with single quantum emitters integrated into their wall could possibly serve as flexible ring resonators for advanced quantum electrodynamic experiments.[4] Furthermore, it was demonstrated that dye liquid can be filled into rolled-up tubes[6]. The ability to combine integrative optical resonators with microfluidics might open interesting options for optical biosensors[7] or lab-on-a-chip devices with new functionalities. However, to apply microtube ring resonators as real devices, Si-based structures operating at room temperature would be highly appreciable due to their compatibility with main stream microelectronics.

It was shown previously that intrinsic SiGe layer stacks[8,9] and single Si layers[10] can roll-up into micro- and nanotubes on Si(001) substrates. In this Letter, we exploit this approach to create well-positioned SiO$_x$/Si microtubes and radial superlattices[11]. The high structural integrity of these objects is revealed in unprecedented detail by targeted focused ion beam preparation and subsequent scanning, cross-sectional and energy filtered transmission electron microscopy (TEM). After thermal annealing, we observe visible light emission at room temperature from the rolled-up microtubes. Optical resonant modes are obtained for free-standing tubes with mode spacings inversely proportional to the tube diameter as expected from Finite-Different-Time-Domain (FDTD) calculations.

We deposit Si layers beyond the critical thickness for plastic relaxation onto 40-70 nm thick Ge sacrificial layers on Si (001) substrates at 300°C by molecular beam epitaxy. Afterwards, SiO$_x$ layers with thicknesses between 5 and 20 nm are thermally evaporated on top of the tensile strained Si layers. To create rolled-up tubes, we use mechanical scratching to define starting edges for an underetching process. The SiO$_x$/Si layers are released from the Si substrate by selectively removing the Ge sacrificial layer by H$_2$O$_2$ at 90ºC. After tube formation, we heated the tubes at 850ºC for 30 min using a rapid thermal annealing process.

Scanning electron microscopy (SEM) was performed to determine the tube diameters. The tube structures were prepared by targeted focused ion beam etching/polishing and investigated by scanning TEM (STEM), cross sectional and energy filtered TEM. STEM was performed by an FEI Dual Beam STRATA400. For TEM and energy filtered TEM, the structures were investigated by a Zeiss LIBRA120 microscope. The optical properties were revealed by fluorescence and micro-photoluminescence (µ-PL) spectroscopy. The excitation source for µ-PL was a double frequency Nd-YVO$_4$ laser operating at 532 nm. The laser light was focused to approximately 2 µm using a 50×microscope objective lens. More details of the setup can be found in Ref 12.

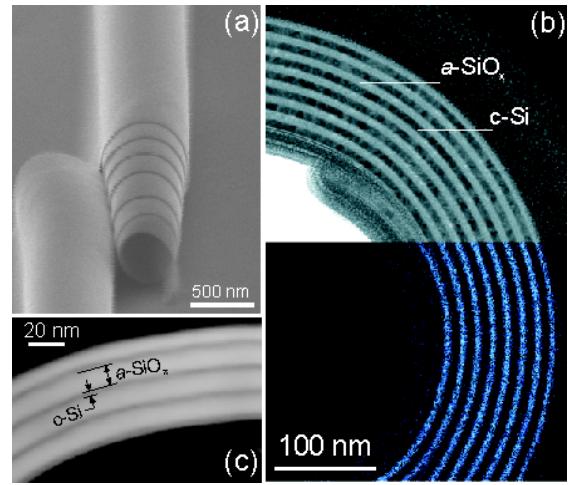

**Fig. 1**: (a) SEM image of a typical inverted rolled up tube from 7 nm SiO$_x$/ 8 nm Si. (b) Cross sectional TEM image (upper part) combined with an oxygen element map (lower part) of a tube wall consisting of 7 nm a-SiO$_x$ (bright area) and 8 nm c-Si (dark area). The blue region in the lower part originates from the area, which contain oxygen. (c) Cross sectional STEM image of a tube wall consisting of 10 nm a-SiO$_x$ (bright area) and 3 nm c-Si (dark area).

Figure 1(a) is an SEM image of a typical rolled-up tube created by 7 nm SiO$_x$/8 nm Si layers. Note that the tensile strain gradient inside the partially relaxed Si layer causes the released SiO$_x$/Si layers to roll downward towards the substrate surface[10]. The average tube diameter is around 450 nm. As shown in this figure, the layer can perform multiple rotations and thus form tubes with multiple-walls. A detailed analysis of a multiple-wall $SiO_x$/Si tube is presented in Fig. 1(b). The upper (lower) part of this figure is a bright field (energy filtered) TEM image. The tube wall consists of eight periods of 7 nm amorphous-$SiO_x$ (bright area) and 8 nm crystalline-Si (dark area), which is created by eight rotations of the $SiO_x$/Si layer. The lower part shows in blue the area which contains oxygen corresponding to the $SiO_x$ regions in the tube walls. Figure 1(c) is an STEM image of a 500 nm-diameter tube with a wall comprising four periods of 10 nm $SiO_x$/3 nm Si. The TEM and the STEM images manifest the structural formation of an *a*-$SiO_x$/*c*-Si *radial superlattice* with layer thicknesses accurately adjustable by MBE growth and thermal evaporation. The number of rotations, i.e. the number of layer periodicities in the wall can be controlled by the underetching time.

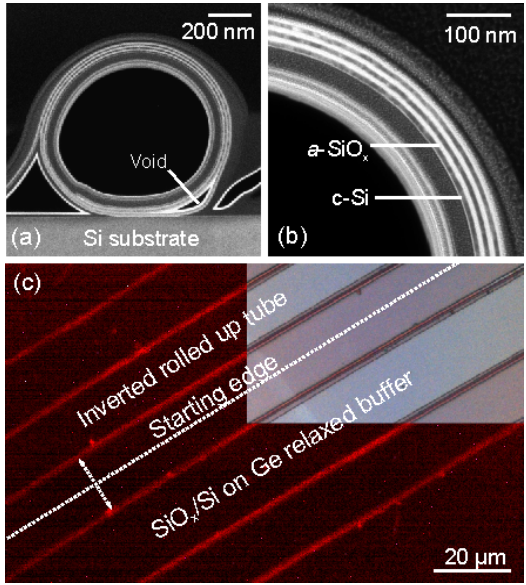

**Fig. 2** (a)-(b) Cross sectional TEM image of a tube and a tube wall consisting of 7 nm a-$SiO_x$ (bright area) and 8 nm c-Si (dark area) after annealing at 850ºC for 30 min. (c) Fluorescence spectroscopy image at room temperature of 10 nm $SiO_x$/8 nm Si tubes. The inset is a light microscopy image.

Figure 2(a)-(b) shows cross sectional TEM images of a $SiO_x$/Si tube and its wall after annealing at 850 ºC for 30 min, respectively. The tube has performed three rotations and consists of a 7nm $SiO_x$/ 8 nm Si bilayer. The wall structure is well-preserved after annealing and there is no evidence of a structural deterioration, implying a high thermal stability of the tube. Apart from a small void in the lower right part of the tube wall (Fig. 2(a)), which occurred during tube formation, sharp interfaces between the Si and $SiO_x$ are revealed. Figure 2(c) shows a fluorescence image at room temperature as well as a light microscopy image (as an inset) of the annealed 10 nm $SiO_x$/8 nm Si tubes. Well-positioned $SiO_x$/Si inverted rolled-up tubes with lengths of more than 100 $\mu$m and uniform visible light emission are displayed. In the case of pure Si tubes annealed under the same conditions, no significant luminescence is observed [13]. We therefore conclude that the light emission originates from the annealed $SiO_x$ area inside the tube walls. At this annealing temperature, we expect that Si nanoclusters form inside an oxide matrix, which gives rise to luminescence in the visible spectral range [14-16]. Moreover, the fluorescence image reveals that the luminescence from the areas, where the $SiO_x$/Si layer is still attached to the Ge sacrificial layer, is significantly less intense compared to areas, where the tubes have formed. In the latter case, the luminescence enhancement is mainly attributed to the number of rotations of the $SiO_x$/Si layer (in this case approximately 3 rotations). By increasing the number of rotations we can locally increase the volume of active $SiO_x$ material compared to the flat regions. However, we also observe a strong intensity enhancement in $SiO_x$/Si microtubes with a single rotation. This observation can be simply explained by efficient light reflection due to an increasing number of semiconductor/air interfaces.

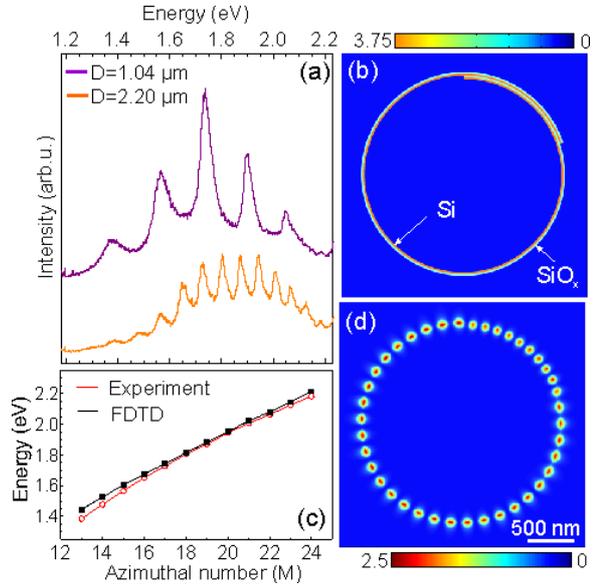

**Fig. 3**: (a) PL spectrum at 8 K of single freestanding $SiO_x$/Si microtubes with a diameter of 1.04 $\mu$m (upper spectrum) and 2.2 $\mu$m (lower spectrum). (b) Refractive index contrast of the rolled-up tube used for the FDTD simulations. The tube wall consists of 20 nm $SiO_x$/20 nm Si. (c) The energy position of the modes extracted from the lower spectrum in (a) compared to the mode energies as a function of the azimuthal number (M) obtained from FDTD simulation. (d) Intensity pattern of the resonant mode at 1.95 eV with M=20 and N=1 from the structure shown in (b).

$\mu$-PL spectroscopy is performed on $SiO_x$/Si tubes at both 8 K and room temperature. No significant shift of the peak position is observed from 8 K to room temperature. In the case of $SiO_x$/Si tubes attached to the Si substrate, a broad PL spectrum is found at around 1.8 eV with a linewidth of about 400 meV (not shown here). The modulation of the PL spectrum appears when the $SiO_x$/Si tubes are disengaged from the substrate. This modulation is ascribed to optical resonant modes occurring inside the tube wall. The PL spectra at 8 K are given in Fig. 3 (a) and reveal pronounced modes with different spacings depending on the tube diameter. The upper spectrum is taken from a 15 nm SiO$_x$/8 nm Si tube with a diameter of 1.04 $\mu$m (sample A). In this case, the SiO$_x$/Si layer performed 3 rotations, as estimated from light microscopy images. The lower spectrum, with narrower mode spacing, corresponds to a tube with a larger diameter of 2.2 $\mu$m and with walls consisting of 1.2 rotations of a 20 nm SiO$_x$/20 nm Si layer (sample B). The mode spacing is inversely proportional to the tube diameter as described elsewhere [5,17].

Finite difference time domain (FDTD) simulations are performed to obtain both the optical resonant energies and the intensity pattern from the rolled up tube structures (Lumerical Solutions software). Modes are labeled according to their azimuthal (M) and radial number (N). Figure 3(b) illustrates the refractive index contrast of sample B, which we use for the simulation. We assume the refractive index of the SiO$_x$ area as a constant value equal to 1.7. The refractive index of the Si area is varied as a function of wavelength [18]. The comparison between the calculated and the measured resonant mode energies (from the lower spectrum in Fig. 3 (a)) as a function of M is presented in Fig. 3(c). The simulated results are in good agreement with the experiment. Figure 3(d) presents the intensity pattern of a mode at an energy of 1.95 eV (M=20 and N=1). This pattern reveals that the light can be confined inside the tube wall despite its small thickness (40 nm). Nevertheless, the limited tube wall thickness is probably the main reason of the line width broadening of the resonant modes. Based on FDTD simulations, we expect that by increasing the number of rotations, the light losses can be reduced, leading to an improvement of the quality factor of the structure and a narrowing of the mode peaks. Moreover, an imperfection of the tubes (such as the voids inside the tube wall) can also deteriorate the quality factor of the structures.

We analyze the polarization state of the emitted light by means of a rotating lambda-half wave plate and a fixed linear polarizer positioned in front of our spectrometer. The PL spectrum shown in Fig. 4 is obtained from a freestanding 20 nm SiO$_x$/ 15 nm Si tube. Figure 4(a) reveals that by rotating the polarization axis with respect to the tube axis we can enhance (upper spectrum) or completely suppress the modes (lower spectrum). In Fig. 4(b) the average peak-to-valley ratio for the modes is shown as a function of the angle between the tube axis and the polarizer axis. These data show that the optical resonant modes observed in the PL spectra are strictly polarized along the tube axis [5], which is caused by repetitive total reflection of the light at the tube walls.

In conclusion, we have released a plastically relaxed Si layer[10] covered with SiO$_x$ to fabricate SiO$_x$/Si radial superlattices and SiO$_x$/Si microtubes. The hybrid SiO$_x$/Si microtubes are thermally stable up to 850 °C and serve as visible light emitters and ring resonators with strictly

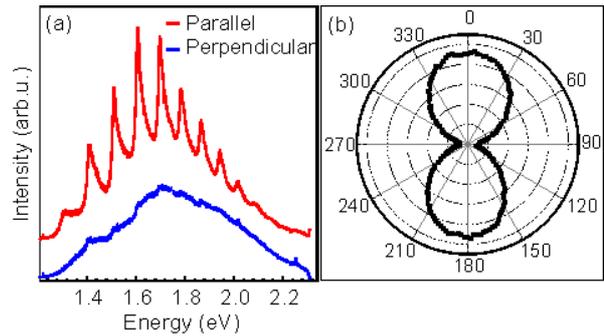

**Fig. 4**: PL measurement at room temperature of a single freestanding SiO$_x$/Si microtube for different polarization configurations. (a) PL spectrum polarized parallel (upper spectrum) and perpendicular (lower spectrum) to the tube axis. (b) Average PL peak-to-valley ratio as a function of the polarization angle with respect to the tube axis. Zero degree represents polarization parallel the tube axis

polarized optical modes at room temperature. Based on this technology, several different functionalities, such as light emission, optical resonances, mechanical flexibility and 2D confined nanofluidic transport can possibly be integrated into a single Si-compatible device.

We thank Y. F. Mei, S. Kiravittaya and Ch. Deneke for fruitful discussions. We are grateful to Carl-Zeiss Company, Oberkochen, Germany for TEM preparation and characterization and FEI Company, Eindhoven, the Netherlands for STEM characterization. The work was financially supported by the BMBF (03N8711).

1 T. J. Kippenberg, S. M. Spillane, D. K. Armani, and K. J. Vahala, Appl. Phys. Lett. **83,** 797 (2003)
2 D. K. Armani, T. J. Kippenberg,, S. M. Spillane, and K. J. Vahala, Nature (London) **421**, 925 (2003)
3 M. L. Gorodetsky, V. S. Ilchenko, H. Mabuchi, E. W.Streed, and H. J. Kimble, Opt. Lett. **20**, 1515 (1995)
4 S. Mendach, R. Songmuang, S. Kiravittaya, A. Rastelli, M. Benyoucef, and O. G. Schmidt, Appl. Phys. Lett. **88**, 111120 (2006)
5 T. Kipp, H. Welsch, Ch. Strelow, Ch. Heyn, and D. Heitmann, Phys. Rev. Lett. **96**, 077403 (2006)
6 Ch. Deneke, N. Y. Jin-Phillipp, I. Loa, and O. G.Schmidt, Appl. Phys. Lett. **84**, 4475 (2004).
7 F. Vollmer, D. Braun, A. Libchaber, M. Khoshsima, I. Teraoka, and S. Arnold, Appl. Phys. Lett. **80** 4057 (2002)
8 O. G. Schmidt and K. Eberl, Nature **410**, 168 (2001).
9 O. G. Schmidt and N. Y. Jin-Phillipp, Appl. Phys.Lett. **78** 3310 (2001)
10 R.Songmuang, Ch. Deneke, and O G. Schmidt. Appl. Phys.Lett. (Accepted)
11 Ch. Deneke and O. G. Schmidt, Appl. Phys. Lett. **85**, 2194 (2004)
12 A. Rastelli, S. Kiravittaya, L. Wang, C. Bauer, O. G. Schmidt, Physica E (Amsterdam) **32**, 29 (2006)
13 Y. Mei private communication
14 H. Rinnert, M. Vergnat, G. Marchal, and A. Burneau, Appl. Phys. Lett. **72**, 3157 (1998).
15 U. Kahler andH. Hofmeister, Opt. Mat. **17**, 83 (2001)
16 L. X. Yi, J. Heitmann, R. Scholz, and M. Zacharias, Appl. Phys. Lett. **81** 4248 (2002)
17 S. L. McCall, A. F. J. Levi, R. E. Slusher, S. J. Pearton, and R. A. Logan, Appl. Phys. Lett. **60**, 289 (1992).
18 The parameters based on Lumerical Solutions software